\documentclass[runningheads]{llncs}
\usepackage[T1]{fontenc}
\usepackage{graphicx}
\usepackage{url}
\usepackage{tabularx}
\usepackage{multirow}
\usepackage{array}
\usepackage{makecell}
\newcolumntype{C}[1]{>{\centering\arraybackslash}m{#1}} 
\usepackage{pifont} 

\newcommand{\cmark}{\ding{51}} 
\newcommand{\xmark}{\ding{55}} 
\begin{document}
\title{ALFA: A Safe-by-Design Approach to Mitigate Quishing Attacks Launched via Fancy QR Codes}

\author{Muhammad Wahid Akram\orcidID{0000-0002-0680-7898} \and
Keshav Sood\orcidID{0000-0002-2127-1438} \and
Muneeb Ul Hassan\orcidID{0000-0001-5109-9547} \and Dhananjay Thiruvady\orcidID{0000-0002-8011-933X}}
\authorrunning{Akram et al.}
%
\institute{School of IT, Deakin University, Geelong, VIC 3217,
Australia
\email{\{s224289198, keshav.sood, dhananjay.thiruvady\}@deakin.edu.au, muneebmh1@gmail.com}}
\maketitle              

\begin{abstract}
Phishing with Quick Response (QR) codes is termed as Quishing. The attackers exploit this method to manipulate individuals into revealing their confidential data. Recently, we see the colorful and fancy representations of QR codes, the 2D matrix of QR codes which does not reflect a typical mixture of black-white modules anymore. Instead, they become more tempting as an attack vector for adversaries which can evade the state-of-the-art deep learning visual-based and other prevailing countermeasures. We introduce "ALFA", a safe-by-design approach, to mitigate Quishing and prevent everyone from accessing the post-scan harmful payload of fancy QR codes. Our method first converts a fancy QR code into the replica of binary grid and then identify the erroneous representation of modules in that grid. Following that, we present "\textit{FAST}" method which can conveniently recover erroneous modules from that binary grid. Afterwards, using this binary grid, our solution extracts the structural features of fancy QR code and predicts its legitimacy using a pre-trained model. The effectiveness of our proposal is demonstrated by the experimental evaluation on a synthetic dataset (containing diverse variations of fancy QR codes) and achieve a FNR of 0.06\% only. We also develop the mobile app to test the practical feasibility of our solution and provide a performance comparison of the app with the real-world QR readers. This comparison further highlights the classification reliability and detection accuracy of this solution in real-world environments.

\keywords{Fancy QR Codes \and Quishing \and Phishing \and QR Code Detection \and QR Code Security}
\end{abstract}
\section{Introduction}
The ever-evolving nature of digitized interaction of individuals are continuously being monitored and exploited by cyber attackers for phishing attempts in numerous ways such as Smishing (SMS Phishing) \cite{nahapetyan2024sms} or Vishing (Voice Phishing) \cite{236328}. Recently, another attack vector is emerged, named as Quishing, in which attackers spread malicious payload using  QR codes. As reported in 2024 by Insikt Group \cite{InsiktGroup2024} and ReliaQuest \cite{ReliaQuest2024} that in overall QR code related attacks, 22\% of them were accounted for Quishing, indicating a 248\% surge as compared to 2023. Moreover, now, Quishing attacks have become more deceptive and persuasive with the use of large language models (LLMs) \cite{akram2025exemplifying}.\par

\begin{figure}
\includegraphics[width=\textwidth]{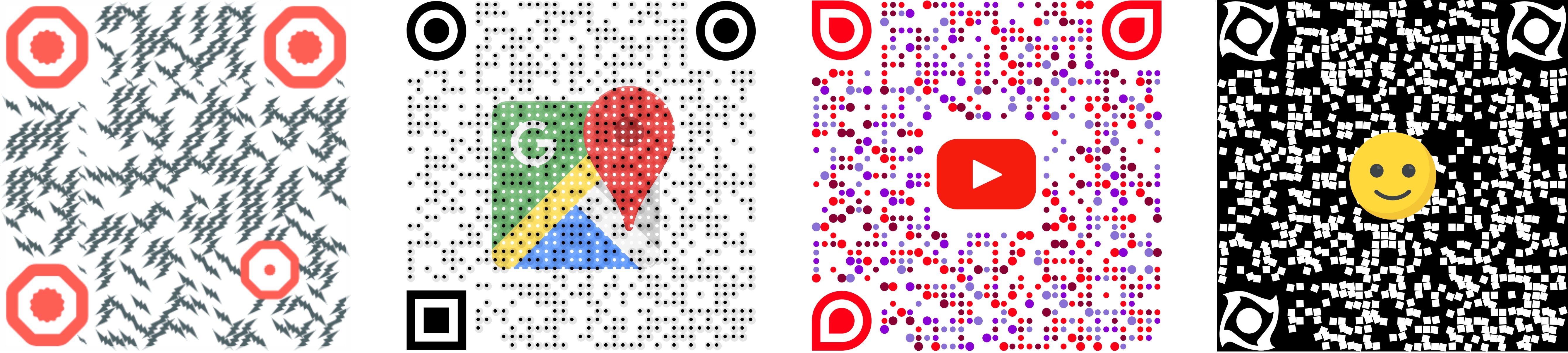}
\caption{Some examples of custom-shaped (fancy) QR codes. They reflect the unusual multi-color illustration of modules, having logos, background image/color.} \label{fig:examples_of_fancy_qr_codes}
\end{figure}

The ordinary looking 2D black-white boxes of QR code matrix can trick anyone, this is because, individuals cannot see the potentially harmful content embedded in it until unless they scan it. Moreover, the recent advancements of custom-shaped (fancy) \cite{QRCodeChimp2025,QRPlanetDesigner2025}, aesthetic \cite{xu2019stylized}, and artistic \cite{xu2021art} QR codes, became more popular among organizations (including banking, education, health, aviation, etc) and individuals for marketing, branding, and other purposes. Usually, artistic and aesthetic QR codes are created by blending an image with black-white QR code where their modules are almost unidentifiable to human eye. In custom-shaped QR codes, modules are much observable even they are of irregular fancy module shapes and colors. They may further come up with background images or colors, foreground icons/logos, etc \cite{QRCodeChimp2025,QRPlanetDesigner2025}, as shown in Fig. \ref{fig:examples_of_fancy_qr_codes}. Therefore, it is to be noted that, this study is emphasized on structural analysis of the observable modules in the custom-shaped QR codes to classify them as phishing and legitimate. In rest of the paper, we referred them as fancy QR codes.\par

There is a wide variety of open-access resources (i.e., QR generators) available that any individual can create fancy QR codes at no cost. Moreover, in this context, the existing studies \cite{su2021artcoder,xu2021art,su2021q,liao2025diffqrcoder}, are mainly focused on proposing effective methods to craft visually pleasing and beautify QR codes and improving their scanning robustness. For instance, authors in \cite{xu2021art} introduced a scanning probability estimation model based on QR code modules. Similarly, Liao et al. \cite{liao2025diffqrcoder} presented DiffQRCoder. A QR code generator based on Diffusion Models. However, a notable aspect lacking in all of these works is they did not consider this 2D binary matrix's exploitation from Quishing angle. This further arises question on the security perspectives of QR code generators.\par

For a QR code, its appearance is the first point of contact for the individuals, therefore, we argue that QR readers should be capable to recognize its legitimacy before accessing the payload which is not possible with URL-driven approaches \cite{tayachi2025quishing,shevchuk2025qr}. Additionally, the visual-based deep learning methods are only trained with black-white QR codes and they also might not be suitable for fancy QR codes which are composed of diverse range of unusual refined patterns \cite{tayachi2025quishing}.\par

By considering the overlooked aspect of identifying the legitimacy of fancy QR codes and shortcomings of existing literature, this paper aiming to fill this gap by proposing a safe-by-design approach named "ALFA", abbreviated as \textbf{A}nti-phishing \textbf{L}ayer for \textbf{F}ancy QR code \textbf{A}ttacks. To better comprehend this concept, we have discussed an example scenario. The fancy QR codes comes with diverse variations unlike standard black-white QR codes. Moreover, the real-world QR readers can effectively decode them in fractions of a second. However, once a QR is scanned, the individuals have no idea of what lies behind it. Leveraging this opacity, adversaries can exploit fancy codes to trigger fraudulent and identity theft practices. Therefore, this vulnerability necessitates a design oriented solution that can practically detect and classify fancy QR codes as phishing or legitimate without looking into the post-scan malicious content. In addition, unlike existing QR readers where they need an external server for processing, this solution can fully operate on the user device, making sure that user's personal details will not be shared or processed with any external server.\par

In recent studies \cite{trad2025detecting,akram2025qr}, authors proposed different methods to structurally analyze QR codes. These solutions are mainly evaluated on the datasets containing particularly standard black-white QR codes. Moreover, authors in \cite{akram2025exemplifying} explicitly mentioned that their solution is not applicable to fancy QR codes. While pixel-based approach in \cite{trad2025detecting} is also not reliable in the scenario of colorful and irregular modules present in fancy codes. In our approach, we first converted a fancy QR into the exact binary grid with the help of identifying the intensity of each of its modules. This grid represents `1' where-ever our algorithm found dark (colorful) modules and `0' for light (white) modules. However, from practical perspective and variation of module shapes in QR codes, it is possible that the binary grid might have erroneous values for 1's and 0's. To recover them, authors in \cite{akram2025qr,trad2025detecting} have no solution, which can further effect their final prediction results in real-world settings. While in our methodology, we presented `\textit{FAST}' method stands for Finder, Alignment, Separators, and Timing patterns. This method is a critical and innovative part of our solution, to recover and correctly label the mislabeled erroneous modules in the initial binary grid according to the standard patterns of the QR codes. We compared the results of our experiments in terms of executing our algorithm with and without \textit{FAST} method. The results showed that our design oriented approach with this method was effective in correctly classifying fancy QR codes with false negative rate (FNR) of 0.06\% only. Additionally, the detection accuracy and practical feasibility of our solution is demonstrated by a smartphone app that we developed using the \textit{flutter} framework, supporting both operating systems i.e., android and iOS.\par

It is worth noting that none of the existing studies have presented any solution for Quishing attacks (in context of fancy QR codes). While our solution with the focus on analyzing the design patterns of fancy codes is a secure and pro-active approach, keeping safe the individuals from accessing potentially harmful content. There are few solutions exists which are relied on mobile webview or sandboxing environment concepts to analyze the QR code's payload before revealing it to users. However, authors in \cite{236356} highlighted that accessing URLs in mobile webview can be dangerous and can trigger other attacks, whereas \cite{kondracki2022droid,nappa2021pow} presented how attackers can evade the android sandboxes and online malware analysis tools based on sandboxing environment. Thereby, the safest option to keep individuals' privacy and providing obstruction against QR-based phishing attacks is by analyze the structural design and we believe that the presented safe-by-design approach is a reliable and lightweight solution for resource-constrained devices like smartphones. Not only that, the devised solution can work along with existing QR readers operative in retail stores, banking systems, or any enterprise security software where hundreds of QR codes scanned on daily basis.\par

The core contributions of our work are as follows:
\begin{itemize}
    \item We presented a secure and design oriented safe-by-design approach "ALFA", for the classification of fancy QR codes against Quishing attacks. Our solution structurally examined the irregular patterns of fancy QR code and converted its modules into the interpretable binary replica which is not possible with url-driven or any other existing visual-based deep learning methods.
    \item We introduced a distinctive method named as `\textit{FAST}', to identify and recover any erroneous mislabeled modules in the binary replica of fancy codes using the standard patterns of QR code. To highlight the mislabeling of modules, we first transformed the binary replica into the regular QR code as image. Later, we illustrated that how and what patterns of QR codes can be correctly restored by our algorithm.
    \item A real-world smartphone app is developed to demonstrate the effectiveness of our solution in practical settings. Moreover, the performance comparison of this app is also provided with the prevailing real-world QR readers available on \textit{Play Store} and \textit{App Store}. This comparison is based on distinct metrics highlighting the detection accuracy of phishing QR codes, resource usage, robustness and latency.
\end{itemize}


\section{Literature Review}
Lately, email has been a leading source of attackers to exploit users' trust and steal their confidential data \cite{desolda2021human}. Furthermore, with the growing adoption of QR codes in email communication as an attachment, attackers can blend multiple phishing techniques together i.e., Quishing and email phishing. Hence, in that case, the malicious payload obscured inside 2D matrix of QR codes can bypass the traditional email security filters that are only able to detect plain-text malicious URLs \cite{Chouinard2021}.\par

While authors in \cite{purwanto2025certifichain} discussed the QR code security in blockchain systems and designed a secure QR-based dual credential verification approach to mitigate forgery and Quishing attacks. Authors in \cite{han2023medusa} showed that how in-app QR code readers in smartphones can be susceptible to MEDUSA attacks where attackers can trigger an in-built function (i.e., \textit{Remotely Accessible Handlers} (RAHs)) of QR readers to initiate the fake payment transactions or can submit session specific authentication tokens. This sort of attacks on QR scanners and worldwide adeptness of QR codes indicate that they are highly vulnerable and appealing target of attackers. Moreover, Zhang et al. \cite{zhang2025demystifying} investigated the QR-based login authentication mechanism and reported six crucial deficiencies that facilitates pathway to various attacks including \textit{Double Login, Authorization Hijacking}. For its security, authors proposed \textit{QRLChecker}. However, this method required developers to create a JSON config file in which they need to pre-define the payload (including the redirection URLs) for the associated QR code.\par

Similarly, Bai et al. \cite{bai2017picking} proposed QR-based secure payment method for POS terminals named as \textit{POSAUTH}. This solution is presented with the aim to overcome the existing attack scenario i.e., \textit{Synchronized Token Lifting and Spending} (STLS), targeting the mobile offline payment systems by verifying the payment token after scanning the QR code. Furthermore, POS-based solution may require a specialized scanner which generally do not support by-default feature to check QR's legitimacy. While attackers may manipulate the QR codes in above scenarios, therefore, it is worth mentioning here that our proposed algorithm can withstand in such scenarios and can also be well-integrated into above discussed solutions which is not possible with any known prevailing QR readers.\par

Few studies \cite{tayachi2025quishing,barron2025quashing,sahu2025optimizing,xue2022screenid} emphasized on mitigating Quishing and QR tampering through URL-driven method, cryptography techniques, and screen dimming feature for the secure authentication of QR-based mobile payment. However, to the best of our knowledge, none of the existing literature considered two primary concepts. Firstly, the utilization of colorful and fancy QR codes for Quishing purposes and secondly, the classification solution of these QR codes by 2D structural feature investigation without accessing the encoded payload. Therefore, this study presented a timely design oriented defensive solution against malicious QR codes to protect individuals falling into the crafty, lucrative, and pervasive attempts of QR-based phishing. Additionally, we summarized the comparison of our work with the state-of-the-art solutions in Table \ref{tab:literature_comparison}.

\begin{table}
\caption{Comparison of Proposed Approach with Existing Solutions (ES)}
\centering
\begin{tabularx}{\textwidth}{
    C{0.85cm}
    |C{5.6cm}
    |C{1.0cm}
    |C{1.0cm}
    |C{1.0cm}
    |C{1.0cm}
    |C{1.0cm}
}
\hline
\textbf{Ref.} & \textbf{ES1} & \textbf{ES2} & \textbf{ES3} & \textbf{ES4} & \textbf{ES5} & \textbf{ES6}\\
\hline
\end{tabularx}
\begin{tabularx}{\textwidth}{
    C{0.85cm}
    p{5.6cm}
    C{1.0cm}
    C{1.0cm}
    C{1.0cm}
    C{1.0cm}
    C{1.0cm}
}
\cite{bai2017picking} & POS token-based solution & \cmark & \xmark & \cmark & \xmark & \cmark \\
\hline
\cite{barron2025quashing} & Cryptography-based solution & \cmark & \xmark & \cmark & \xmark & \cmark \\
\hline
\cite{purwanto2025certifichain} & Digital credentials based solution & \xmark & \xmark & \xmark & \xmark & \xmark \\
\hline
\cite{sahu2025optimizing} & Cryptography-based solution  & \cmark & \xmark & \cmark & \xmark & \cmark \\
\hline
\cite{tayachi2025quishing} & URL-driven method  & \cmark & \xmark & \cmark & \cmark & \cmark \\
\hline
\cite{xue2022screenid} & Mobile dimming feature based solution& \cmark & \xmark & \cmark & \xmark & \cmark \\
\hline
\cite{zhang2025demystifying} & Rule-based testing approach  & \xmark & \xmark & \xmark & \xmark & \xmark \\
\hline
\textbf{Ours} & QR structural based approach & \cmark & \cmark & \cmark & \cmark & \cmark \\
\hline
\end{tabularx}
\label{tab:literature_comparison}
\vspace{1ex}
\parbox{1\linewidth}{\centering\footnotesize
\textbf{ES1:} \textit{What method or technique authors presented to address the research problem?}, \textbf{ES2:} \textit{Does the study targeting Quishing aspect of QR codes?}, \textbf{ES3:} \textit{Does the study considered the structural analysis of QR image?}, \textbf{ES4:} \textit{Does the solution applicable to real-world printed QR codes?}, \textbf{ES5:} \textit{Does the solution independent of a dedicated scanner?}, \textbf{ES6:} \textit{Does the solution evaluated with real-world Quishing dataset or scenarios?}, (\xmark): \textit{No}, (\cmark): \textit{Yes}.}
\end{table}

\section{Preliminaries of ALFA}

\subsection{Difference Between Fancy and Black-White QR Codes}
\begin{figure}
\includegraphics[width=\textwidth]{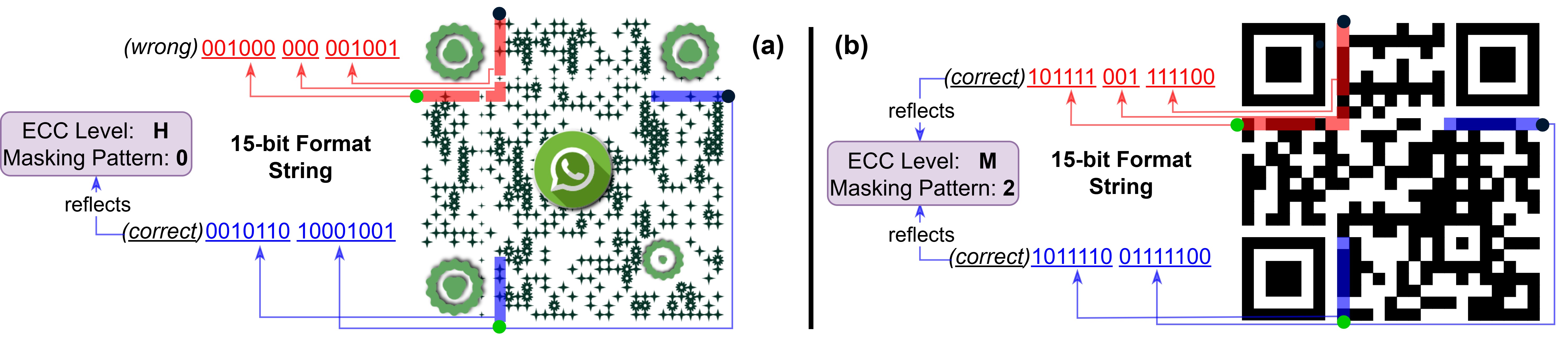}
\caption{Demonstration of differences between a fancy and black-white QR code. Green and black dots show the starting and ending points of reading format bits, respectively.} \label{fig:diff_bw_fancy_and_bw}
\end{figure}

A QR code whether fancy or black-white, is build with standard properties which are enclosed in it at the time of its generation. This includes finder, alignment, and timing patterns, separators, quiet zone, format bits (contains Error Correction Codes (ECC) level and masking pattern), version bits, remainder bits, error correction codes and data modules. Also, existing literature has already discussed them in detail \cite{krombholz2014qr,akram2025qr}, therefore, readers are advised to check the cited references to get more familiar with anatomy of QR codes. A key comparison of fancy and black-white QR codes is presented in Table \ref{tab:diff_bw_fancy_and_bw}. Moreover, below we have discussed their fundamental differences which are also shown in Fig. \ref{fig:diff_bw_fancy_and_bw}.\par

\begin{table}
\caption{Key Comparison Between Fancy and Black-White QR Codes}
\centering
\begin{tabularx}{\textwidth}{
    C{5.50cm}
    |C{5.50cm}
}
\hline
\textbf{Fancy QR Codes} & \textbf{Black-White QR Codes}\\
\hline
\end{tabularx}
\begin{tabularx}{\textwidth}{
    p{5.50cm}
    |p{5.50cm}
}
Having fancy shapes, logos, etc & Always black modules on white layout \\
Irregular refined patterns & Square modules only \\
High dependence on ECC-levels & Standard use of ECC-levels \\
Scan-ability compromised under noise \& blur effects & Maintain robustness under noise \& blur conditions \\
Structural analysis is more challenging & Relatively simple and easy \\
Hard to precisely localize & Easy to perform detection \& cropping \\
High risk of visual manipulation & Low risk of visual manipulation\\

\hline
\end{tabularx}
\label{tab:diff_bw_fancy_and_bw}
\end{table}

\subsubsection{Modules Representation}
The fancy QR code modules are customized with stylish and unusual patterns to make them more appealing. One of the example is shown in Fig. \ref{fig:diff_bw_fancy_and_bw}. Notably, both QR codes contained the same data. However, the combination of multiple colors, authentic logo (in center), circular shape of finder and alignment patterns, and irregular size of modules in Fig. \ref{fig:diff_bw_fancy_and_bw} (a), makes the actual difference in the presentation as compared to Fig. \ref{fig:diff_bw_fancy_and_bw} (b). Undoubtedly, Fig. \ref{fig:diff_bw_fancy_and_bw} (a) is engaging for individuals and this is exact reason what attackers can exploit to mislead and deceive them into disclose the confidential details.

\subsubsection{Position of Format Information Bits}
Generally, format bits appear in every QR code in two positions. From Fig. \ref{fig:diff_bw_fancy_and_bw} (b), black-white QR code reveals that first occurrence of format bits is in \textit{top-left} position and second one is split in two places starting from \textit{bottom-left} and ending in \textit{top-right}. Moreover, from Fig. \ref{fig:diff_bw_fancy_and_bw} (a), we read the formats bits of fancy QR and found that the \textit{top-left} combination of bits are not correct as we verified them with the standard possible format strings \cite{thonky2023}. While, second occurrence of bits are found in \cite{thonky2023}. From this, we concluded that it is not necessary for the scanners to correctly read the format bits from both occurrences. Due to which, fancy QR codes have another advantage of utilized format bits modules for unusual patterns.

\subsubsection{Error Correction Codes (ECC) Reliance}
Because of this features, partially damaged QR codes upto 30\% (level H (High)) can still scannable by the QR readers \cite{krombholz2014qr}. From Fig. \ref{fig:diff_bw_fancy_and_bw}, both QR codes with same data have different ECC levels. Like, fancy QR has \textbf{ECC-level H (High)}, where black-white QR has \textbf{ECC-level M (Medium)}. Economically, ECC-level does not affect at all, however, if QR like Fig. \ref{fig:diff_bw_fancy_and_bw} (a) is already enclosed with ECC-level H then its scan-ability can be more sacrificed in presence of further distortion or low-quality printing.

\section{ALFA: Proposed System Design and Methodology}
Due to the enticing and diverse nature of fancy codes, they can deceptively evade the former efforts such as url-driven and visual based deep learning techniques. Moreover, we found that the relevant prevailing literature lacked the consideration of proposing solutions for mitigating Quishing in context of fancy QR codes. Therefore, to overcome these shortcoming, we presented the secure design oriented approach named "ALFA", for quashing Quishing of fancy QR codes. Our solution is focused on interpreting the anatomy of fancy QR to detect its legitimacy without extracting the potentially harmful payload. We also presented the \textit{FAST} method which can reliably recover and produce the binary replica of fancy codes, through which, later, we extracted and computed the structural features. Furthermore, we fed these features into the pre-trained ML model (described later in this section) for prediction. Now, before diving into the operational workflow of the proposed system, we first discussed the process of dataset preparation for the experimental evaluation of our study.

\subsection{Formation of Fancy QR Codes Dataset}
Based on the literature, no relevant dataset is available that highlights the properties and variations of fancy QR codes as described earlier. Considering this, we prepared our own synthetic dataset. To be more generalize, we utilized 11 publicly available QR generators supporting fancy patterns. To differentiate the phishing and legitimate QR codes, we utilized an existing URLs dataset \cite{prasad2024phiusiil}. We randomly selected 100 URLs (50 for phishing and 50 for legitimate). Then, we provided these URLs to QR generators which output the QR code with their default settings including the size (\textit{height, width}) and image format (\textit{.png}).

\subsection{Operational Working of Devised Methodology}
\begin{figure}
\includegraphics[width=\textwidth]{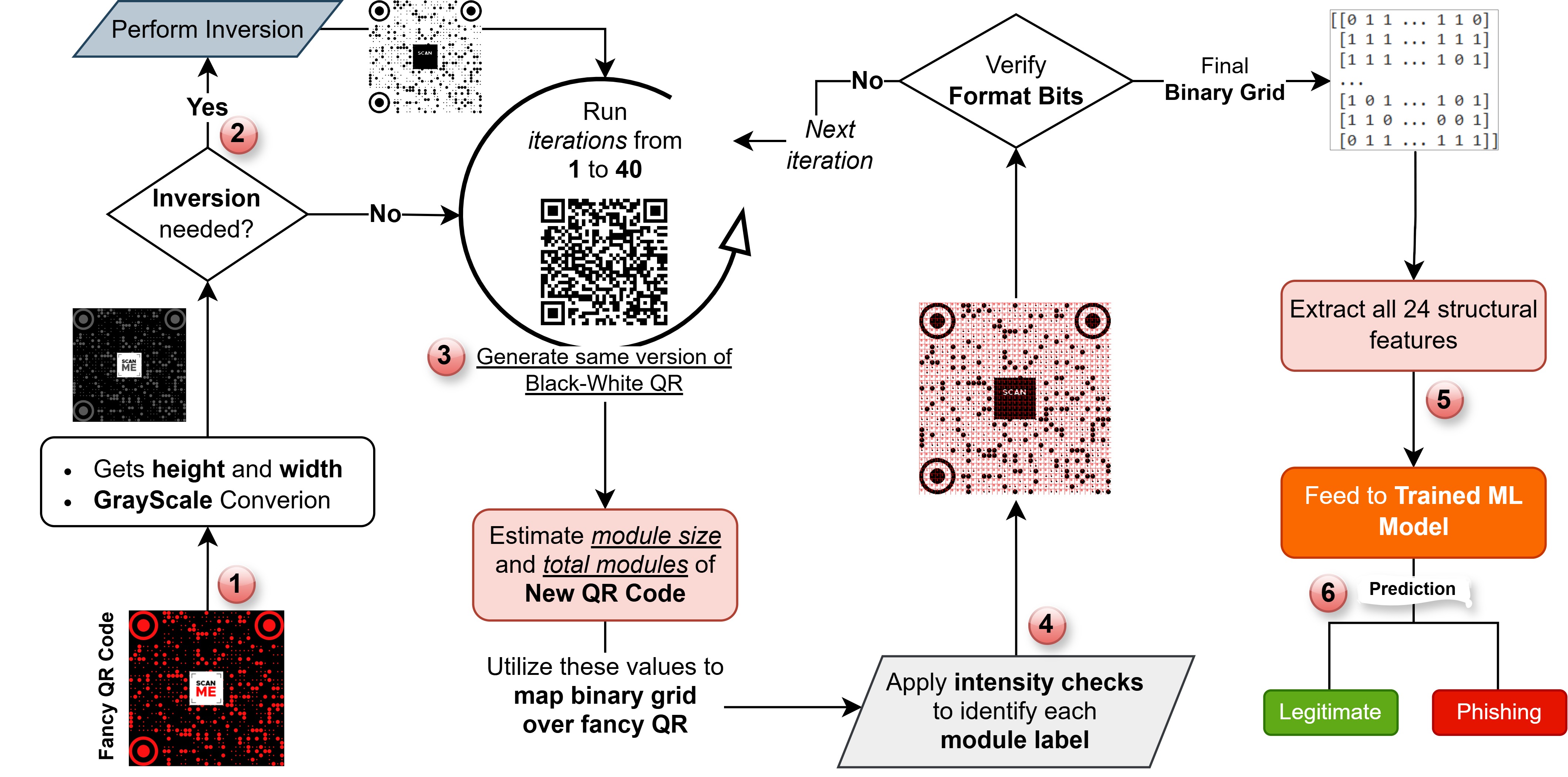}
\caption{The proposed methodology is composed of series of 6 steps. Initiated from inputting and acquiring visual properties of fancy QR. Later, inversion operation is performed and then, an iteration of version 1 to 40 executed to form binary replica, through which, 24 structural features were extracted and fed into model for prediction.} \label{fig:fancy_quishing_methodology}
\end{figure}

The devised methodology is composed of series of steps as shown in Fig. \ref{fig:fancy_quishing_methodology}. For clearer understanding, we have individually outlined each step as follows.

\begin{enumerate}
    \item \textbf{Identify size and grayscale conversion:} Firstly, a fancy QR code is provided to the algorithm which identified its size (\textit{height, width}) and further converted it into a grayscale image.
    
    \item \textbf{Inversion check:} Our algorithm spot white modules in the QR code as `0' and black as `1'. But this is possible that fancy QR codes might have different background color other than white. Therefore, in this step, we identified whether the layout of given QR is inverted (meaning that there are light modules in the QR code with a dark background) or not. In case of this, we computed the ratio of black and white pixels by applying the OSTU's thresholding. Generally, a standard QR have white pixels in majority than black pixels, whereas the opposite scenario might be possible with inverted QR. Following this, our algorithm performed the binary inversion if the ratio of white pixels is below than 51\%. By doing so, the final fancy QR is restored to the correct black-white distribution as expected by our algorithm.

    \item \textbf{Generate new black-white QR code:} In this step, a similar version of black-white QR is generated as of fancy QR. However, the questions arisen that \textit{how} algorithm identified the correct version of fancy QR and \textit{why} black-white QR is required. We first addressed \textit{how}. For this, we know that there are exactly 40 standard versions of QR codes. So, our algorithm starts an iteration of version 1 to 40 and a new black-white QR generated at every iteration. During this, we evaluated the values (of module size, total modules) and a binary grid is also produced for the fancy QR code accordingly. After that, the algorithm extracted and verified the format bits from binary grid with \cite{thonky2023}. If verified, the current iteration is considered as exact version of fancy QR and subsequent steps are performed sequentially. Otherwise, until 40 iterations, if none of the version verified or matched with fancy QR then algorithm gave an output as \textit{``cannot identify fancy QR code"}. Next, \textit{why} we need to identify the exact version of fancy QR? As discussed in section 3.2, authors in \cite{akram2025qr} are dependent on estimating the module size of black-white QR. But in case of fancy QR, because of the irregular modules representation we cannot estimate or fix the size of modules. However, if we have a black-white QR of same version and size as of fancy QR, then we can apply the module size estimation technique on newly generated black-white QR. This will output the module size and total modules of it. Utilizing these values, in our approach, we further produced the binary grid which later mapped over the fancy QR code for format bits verification. Thereby, we then extracted and computed the protocol-level and statistical features.
    
    \item \textbf{Intensity checks for binary grid:} 
    Here, we described that how the algorithm find the intensity of each module in fancy QR and formed the final binary grid. The default range of each module is 0 (black) to 255 (white). In our algorithm, if the mean intensity of each module area is < 189 then it is considered as black (1) module Otherwise, if mean intensity is >= 189 then the module is labeled as `0' (white). This is the standard module identification method we defined for our algorithm. However, during our experiment, there were few fancy QR codes whose individual modules representation were too small. In that case, the mean intensity with 189 threshold was not successful. Therefore, we defined a second check for small modules if they are having mean intensity < 238 then it labeled as black (1) and white (0) otherwise. Now, to identify when to apply 189 or 238 threshold, the fancy QR first passed with 189 mean intensity check and computed the total number of 1's and 0's in it using the binary grid. Then we check, if 0's percentage is > 65 in the QR. If it is, then we again produced the new binary grid for fancy QR with the mean intensity 238. Because, in case of small modules and with mean intensity 189, we verified the correct binary grid by extracting and matching the format bits with \cite{thonky2023}. We noticed that the QR modules were there but the algorithm was unable to label it as 1. In this way, finally we ended up defining two checks for evaluating mean intensity of each module.

    \item \textbf{Features extraction:} 
    Next, the algorithm extracted the 24 structural features of fancy QR code using the binary grid including the protocol-level and statistical features according to the technique employed in \cite{akram2025qr}.

    \item \textbf{Prediction:} Lastly, the extracted features are fed into the pre-trained ML model (i.e., XGBoost) to predict the legitimacy of fancy QR code as phishing or legitimate.
\end{enumerate}

 \subsection{Conversion of Binary Grid to Black-White QR Code}
\begin{figure}
\includegraphics[width=\textwidth]{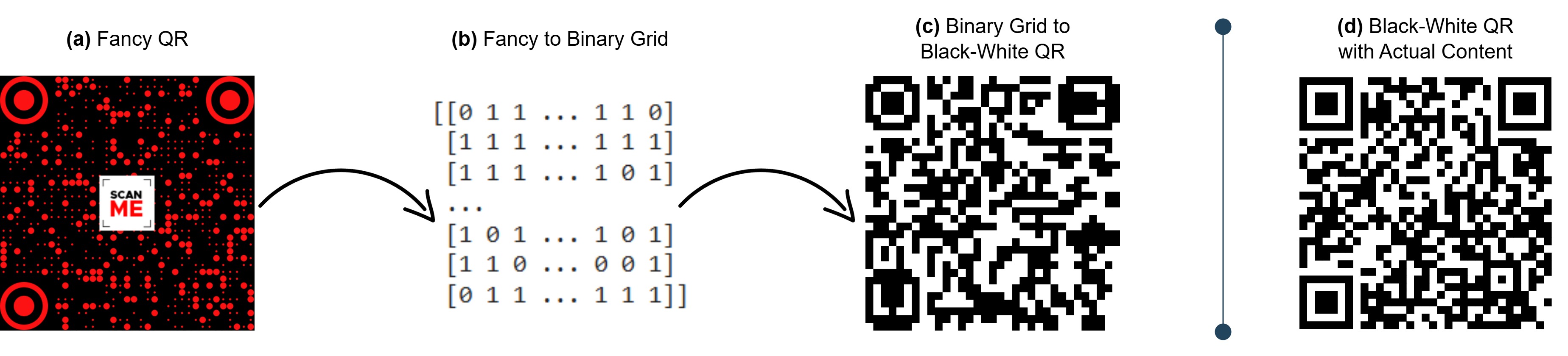}
\caption{The conversion from fancy code to binary replica and to black-white QR code is shown in (a) to (c). In (d), the actual black-white QR encoded with URL of fancy QR is also presented to visualize the erroneous mislabeled modules in the binary grid.} \label{fig:conversion_of_binary_to_bw_qr}
\end{figure}

 From above working of the algorithm, we demonstrated that how we successfully extracted the structural features of fancy QR code with the verified binary grid. Apart from this, we noticed that the produced binary grid in step 4 is not the exact replica of each module in the fancy QR code. To ensure this, we created a black-white QR code of the binary grid of fancy QR as shown in Fig. \ref{fig:conversion_of_binary_to_bw_qr}. In this figure, (a) represents the fancy QR itself, (b) is the binary grid replica, (c) is the reflected black-white QR of binary grid, and (d) is the standard black-white QR code with the actual URL stored in the fancy QR code. Note that, (c) is the representation of binary grid on the base of which the final prediction is made by the model. Moreover, the comparison of (c) and (d) showed the notable difference. This implied that although our algorithm verified the format bits in the binary grid, however, there are other regions (specifically if we say ``modules") that are not correctly labeled as `1' (black) or `0' (white). The differences in Fig. \ref{fig:conversion_of_binary_to_bw_qr} (c) found in the areas of fancy QR codes where there is an image in the background, mislabeling of finder and alignment patterns due to their unusual shapes, and mislabeling of timing patterns as well.  Because of which, there is a possibility of misclassification by the trained model.\par

To avoid the potential misclassification, we further observed that we can recover some of mislabeled modules but not all of them. The recoverable regions in Fig. \ref{fig:conversion_of_binary_to_bw_qr} (c) includes \textbf{f}inder patterns, \textbf{a}lignment patterns, \textbf{s}eparators, and \textbf{t}iming patterns. Following this, we introduced \textit{FAST} method which can be applied in the step 4 of our methodology where the binary replica of fancy code is generated. Other than that, specifically the data part modules can only be recovered in case, if the proposed algorithm knows the actual URL. While our classification approach is all-in-all based on predicting and mitigation of QR code phishing without having the need to look into potential harmful payload. Therefore, we are not focused on recovering these modules. Furthermore, below we have presented the \textit{FAST} method, utilized to recover the mislabeled modules in Fig. \ref{fig:conversion_of_binary_to_bw_qr} (c).

\subsubsection{\textit{FAST}: Recovery of Mislabeled Modules}
This method is stand for \textbf{f}inders, \textbf{a}lignment, \textbf{s}eparators, and \textbf{t}iming patterns. These patterns have a specific fixed position in every fancy or black-white QR code. We read modules of each pattern one-by-one and analyzed their structure according to the standard pattern. If any module in the region is mislabeled as `0' then our \textit{FAST} method corrected it to `1' and vice versa. Additionally, the method keep updated the binary grid as well after examining all four regions. The individual recovery process of each pattern is described as follows and one of the example is presented in Fig. \ref{fig:recovery_of_mislabeled_modules}:
 
\begin{figure}
\includegraphics[width=\textwidth]{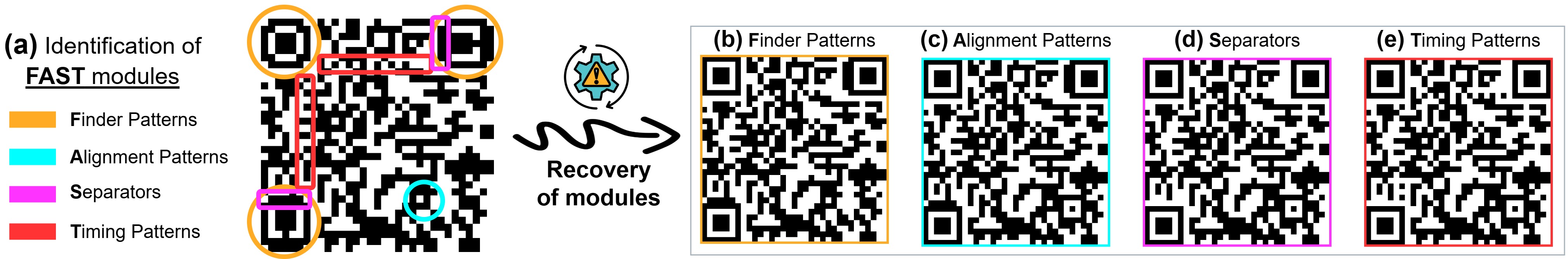}
\caption{The identification of recoverable patterns from binary grid is shown in (a). While (b) to (e) highlight the correction of patterns after applying the \textit{FAST} method.} \label{fig:recovery_of_mislabeled_modules}
\end{figure}

 \begin{itemize}
     \item \textbf{Finder patterns:} are visible on 3 corners of a QR code except the \textit{bottom-right} corner. From Fig. \ref{fig:recovery_of_mislabeled_modules} (a), we can see that many modules in finder patterns are mislabeled as `0' instead of `1' and `1' where it should be `0'. In general, these patterns are matrix of 7x7 in every QR code. The outermost ring should always be labeled as black modules, then next inner ring 5x5 is of white modules, and lastly a black square of 3x3 matrix standardly visible in the center of the finder patterns. Following this standard, we recovered and fixed these patterns in every binary grid of fancy QR code with one example as shown in Fig. \ref{fig:recovery_of_mislabeled_modules} (b).
     
     \item \textbf{Alignment patterns:} have pre-defined specific locations\footnote{https://www.thonky.com/qr-code-tutorial/alignment-pattern-locations} in every QR code except version 1 QR. Structurally, an alignment pattern is a matrix of 5x5 modules. The outermost ring is of black modules, next white inner ring with 3x3 matrix,
     is represented as in white, 
     and lastly 1 black module at the center. From Fig. \ref{fig:recovery_of_mislabeled_modules} (a), we can observe that the alignment pattern is not correctly labeled by the algorithm. However, the \textit{FAST} method corrected all of its mislabeled modules and update the binary grid as well in Fig. \ref{fig:recovery_of_mislabeled_modules} (c).
     
     \item \textbf{Separators:} are always of 1 width white module in size and positioned around finder patterns to separate them from other modules. In Fig. \ref{fig:recovery_of_mislabeled_modules} (a), the separators were mislabeded around top-right and bottom-left of the QR as highlighted with \textit{pink} color. However, the successful recovery of these modules by converting them from `1' to `0' is shown in Fig. \ref{fig:recovery_of_mislabeled_modules} (d).
     
     \item \textbf{Timing patterns:} are visible in a sequence of black-white-black (1-0-1) modules in a QR code along the vertical and horizontal axis. They always assigned the fixed position on 6th row and 6th column, horizontally and vertically in a QR code, respectively. Following Fig. \ref{fig:recovery_of_mislabeled_modules} (a), our \textit{FAST} method looked for the standard positions of timing patterns in the binary grid to first identify the labeling errors and then perform the recovery operation. An example of revised binary grid is manifested in Fig. \ref{fig:recovery_of_mislabeled_modules} (e).
     
 \end{itemize}

\subsection{Pre-trained Classification Model}
As our approach is emphasized on structural features of fancy QR codes so instead of training a new model for classification we have employed the machine learning (ML) model (i.e., XGBoost) trained by \cite{akram2025qr}. Authors individually extracted from 400,000 black-white QR codes dataset. Moreover, for the model evaluation, they employed Optuna with objective function including 5-fold stratified cross-validation. Also, authors claimed the satisfactory results with various evaluation metrics. More importantly, no such relevant study or pre-trained model is existed in prior literature, thus, we opted to employ author's model for the prediction of fancy QR codes in our study. In addition, it is essential to highlight that we extracted the same 24 structural features from fancy QR codes which is identical to the features considered by authors for training purposes.\par

One other critical concern is, their approach cannot be directly applicable to fancy codes at all as authors have only designed it for black-white QR codes. Even if we converted the fancy QRs to gray ones and apply some thresholding methods, still it will not work. This is because of the unique approach presented by the authors, where they estimated the module size in the black-white QR codes and subsequently extracted features. As fancy QR codes comes with irregular module sizes so author's work is unlikely to directly apply in our case.

 \section{Classification Results}
We experimented the proposed approach on 100 fancy QR codes dataset using Google Colab. Our script was written in python language with the supported libraries including \textit{cv2, pyzbar, Image, segno}. Using the \textit{for} loop iteration, each QR code from the dataset is passed through the steps illustrated in the devised methodology. Subsequently, all the extracted structural features and prediction made by the pre-trained model were stored in a \textit{csv} file.\par
We ran two different experiments. The first experiment is carried out by following the steps presented in Fig. \ref{fig:fancy_quishing_methodology} without recovering the mislabeling of modules in the binary grid. However, in second experiment, we utilized the \textit{FAST} technique to recover the mislabeled modules first and then made the prediction accordingly. By doing so, we presented the difference of the predicted results during both experiments in Table \ref{tab:classification_results}.\par

\begin{table}
\caption{Classification Results}
\centering
\begin{tabularx}{\textwidth}{
    C{2.00cm}
    |C{1.50cm}
    |C{2.75cm}
    |C{2.85cm}
    |C{2.50cm}
}
\hline
\textbf{Category} & \textbf{Total} & \textbf{Correct Predictions} & \textbf{Wrong Predictions} & \textbf{Unsuccessful Scans}\\
\hline
\end{tabularx}
\begin{tabularx}{\textwidth}{
    C{11.50cm}
}
\rule{0pt}{1.2em}
\textbf{(Without FAST Method)} \\
\hline
\end{tabularx}
\begin{tabularx}{\textwidth}{
    C{2.00cm}
    C{1.50cm}
    C{2.75cm}
    C{2.85cm}
    C{2.50cm}
}
Phishing & 50 & 44 & 3 & 3 \\
Legitimate & 50 & 15 & 34 & 1 \\
\hline
\end{tabularx}
\begin{tabularx}{\textwidth}{
    C{11.50cm}
}
\rule{0pt}{1.2em}
\textbf{(With FAST Method)} \\
\hline
\end{tabularx}
\begin{tabularx}{\textwidth}{
    C{2.00cm}
    C{1.50cm}
    C{2.75cm}
    C{2.85cm}
    C{2.50cm}
}
Phishing & 50 & 44 & 3 & 3 \\
Legitimate & 50 & 22 & 27 & 1 \\
\hline
\end{tabularx}
\label{tab:classification_results}
\end{table}

In this table, phishing QR codes are considered as positive class. From first part of the table (without \textit{FAST} method), the pre-trained ML model correctly predicted 44 phishing samples as phishing whereas only 3 out of 50 phishing QR codes are classified as legitimate. This outcome resulted in a false negative rate (FNR) of only 0.06\% approximately. However, there are 3 QR codes for which the algorithm failed to extract any structural features. Due to which, it resulted into unsuccessful prediction. In addition, for 50 legitimate QR codes, the model correctly classified 15 samples as legitimate and 34 of them are misclassified as phishing QR codes with a false positive rate (FPR) around 0.7\%. While only 1 legitimate sample was unclassified by the model.\par
From second part of the table (where \textit{FAST} method is applied), the presented results showed a moderate improvement against FPR. For phishing samples, the predicted results were same as compared to the first part of the table. While the correctly predicted legitimate samples are significantly increased to 22 with lesser incorrect predictions 27 and 1 unsuccessful results. If we compare the results from Table \ref{tab:classification_results}, it is evident that the performance of proposed algorithm with \textit{FAST} method proved as aid in successful extraction of structural features of fancy QR codes. Also, the notable reduced rates of FNR and FPR demonstrated the classification reliability and detection sensitivity of our approach.

\subsection{Comparative Analysis}
This study presented a pioneer design oriented approach to mitigate Quishing attacks in context of fancy QR codes. Moreover, to the best of our knowledge and literature review, no existing study has considered relevant scenario. Due to this rationale, the possibility of demonstrating a meaningful and fair comparison with prior studies becomes limited. As highlighted in literature review section, many studies investigated various QR-related security areas and more specifically in phishing domain as well. However, a direct comparison between our approach and authors' approaches is not feasible. The existing studies are either relied on URL-driven or visual-based deep learning methods. In contrast, our study is emphasized on mitigating Quishing by examining the unusual refined patterns of fancy codes which has not been explored before. Besides that, we practically implemented our solution through a smartphone app to manifest its real-world feasibility. Moreover, there are plenty of android and iOS apps available in \textit{Play Store} and \textit{App Store} to alleviate Quishing. But their methodology is based on URL-driven techniques. Despite that, we considered other distinct metrics to perform the comparative evaluation of our app with the real-word QR readers. The detailed discussion of this comparison is presented in the next section.

\section{Practical Evaluation of Our Solution}
In this section, we demonstrated the practical feasibility of our proposed method through a smartphone app. For evaluation, we performed the comparative analysis of our app with the real-world iOS and android apps against distinct metrics. Our app is developed using \textit{flutter} framework supporting both android and iOS platforms. Moreover, we considered the android version of our app for the experiment. We selected 10 phishing and 10 legitimate fancy QR codes from our dataset. The results from this overall app analysis are presented in Table \ref{tab:feasibility_assessment}.\par

\begin{table}
\caption{Practical Evaluation of Mobile App}
\centering
\begin{tabularx}{\textwidth}{
    C{0.50cm}
    |C{1.90cm}
    |C{1.90cm}
    |C{1.90cm}
    |C{2.20cm}
    |C{1.50cm}
    |C{1.50cm}
}
\hline
\textbf{OS} & \textbf{App Names*} & \textbf{Classifying Method} & \textbf{Correct Predictions} & \textbf{Wrong \& Unsuccessful} & \textbf{Runtime (s)} & \textbf{App Size (MB)} \\
\hline
\end{tabularx}
\renewcommand{\arraystretch}{1.5}
\begin{tabularx}{\textwidth}{
    C{0.50cm}
    |C{1.90cm}
    C{1.90cm}
    C{1.90cm}
    C{2.20cm}
    C{1.50cm}
    C{1.50cm}
}

\multirow{2}{*}{\raisebox{-3\height}{\rotatebox[origin=c]{90}{\makecell{iOS}}}} & QR Safe Scanner & URL-based & L: 10/10\ \ \ \ \ \ \ P: 10/10 & L: 0(w), 0(u)\ \ \ P: 0(w), 0(u) & 2.913 & 33.7 \\
 & QR Trust by Qerberos & URL-based & L: 7/10\ \ \ \ \ \ \ P: 4/10 & L: 0(w), 3(u)\ \ \ P: 1(w), 5(u) & 4.624 & 4.5 \\
 & QRSafeScan & URL-based & L: 10/10\ \ \ \ \ \ \ P: 4/10 & L: 0(w), 0(u)\ \ \ P: 6(w), 0(u) & 13.879 & 16.5 \\
\hline
\multirow{2}{*}{\raisebox{-2.2\height}{\rotatebox[origin=c]{90}{\makecell{Android}}}} & QRDefender & URL-based & L: 10/10\ \ \ \ \ \ \ P: 5/10 & L: 0(w), 0(u)\ \ \ P: 5(w), 0(u) & 2.327 & 36.08 \\
& QR Scanner & URL-based & L: 10/10\ \ \ \ \ \ \ P: 4/10 & L: 0(w), 0(u)\ \ \ P: 5(w), 1(u) & 2.924 & 39.84 \\
& QR Scanner Trend Micro & URL-based & L: 8/10\ \ \ \ \ \ \ P: 2/10 & L: 0(w), 2(u)\ \ \ P: 2(w), 6(u) & 1.998 & 26.76 \\
& \textbf{ALFA (Ours)} & Structural based & L: \textbf{6/10}\ \ \ \ \ \ \ P: \textbf{9/10} & L: \textbf{4}(w), \textbf{0}(u)\ \ \ P: \textbf{0}(w), \textbf{1}(u) & \textbf{3.421} & 72.50 \\
\hline

\end{tabularx}
\label{tab:feasibility_assessment}
\parbox{1\linewidth}{\centering\footnotesize*The benchmark apps can be accessed via links provided in github repository.}
\end{table}

The table highlighted that existing apps are based on url-driven methods, while our study presented a QR-based structural approach to mitigate Quishing attacks. Furthermore, out of 10 legitimate (L) fancy QR codes, our app correctly predicted 6 of them as legitimate, while 4 of them was wrongly (w) predicted as phishing. For 10 phishing (P) fancy codes, the correct outcome was 9 out of 10 whereas for only 1 code the app was unable to extract the structural features which leads to unsuccessful (u) result. Moreover, the average runtime of our app for prediction after scanning was just 3.421 seconds which is almost similar or better than the prior apps. While the storage size of our app is bigger than the others. This is only because of the app development with hybrid framework (i.e., \textit{flutter}) which generally acquired more storage size than the native development.

Overall, Table \ref{tab:feasibility_assessment} demonstrated that the performance of our app in terms of Quishing detection, resource usage, robustness, and latency is satisfactory. From comparison, it also indicated that the structural-based analysis for the detection of phishing QR codes is a pro-active method which can protects users from accessing the post-scan malicious content. Although, there are false positives in our results, however, this aspect cannot undermine the actual objective of our study which we have demonstrated, is to identify phishing intention of the adversaries through a safe-by-design approach. Also, with our solution "ALFA", the exploitation of fancy representation of QR codes by attackers can be effectively detected and mitigated in real-world environment. Moreover, our proposed method is a ready-made solution that can practically work along not only with existing QR scanners but can also be integrated within retail systems, security softwares, or banking apps. In addition, the utilized dataset of fancy QR codes in our experiments and the screenshots of our app results can be found on our publicly available github repository\footnote{https://github.com/mwahid905/alfa-quishing-detection-framework}.

\section{Findings and Further Discussion}
Our study aimed to classify the fancy QR codes that can be exploited by the attackers to employ for illicit objectives. The existing literature is emphasized on introducing effective methods to produce fancy, aesthetic, and artistic QR codes. However, the stylish look of QR codes may compromise the scanning process. Therefore, few studies focused on improving the scanning robustness of the QR readers. While the Quishing aspect of fancy QR codes is remained underexplored in the prior studies. Therefore, we are the first one introducing a structural features based approach "ALFA", by which QR readers can be able to detect phishing objective of fancy QR codes hidden behind the fancy representation. Authors in \cite{akram2025qr}, proposed an approach for black-white QRs classification, however, their approach is not applicable to fancy QRs.\par
The structural features can further provide the interpretation details of the classification method which is not possible with pixel-based and deep learning based approaches. Moreover, the experimental results are the evidence which demonstrated the effectiveness of our approach in correctly classifying phishing QR codes whether it is composed of fancy layout patterns or black-white modules. Additionally, the practical feasibility of our approach is highlighted by the developed the mobile app which further shows that this approach is resource-efficient and also can work along with the existing QR readers.\par
Besides that, the presented approach can be further improved in few areas. For example, to produce the binary grid for the fancy QR, we defined two different intensity checks (238 threshold for tiny modules and 189 for other size of modules). Although, we have a dynamic dataset of fancy QR codes, however, these conditions may not work for all type of fancy QR codes in the wild. Another challenge we encountered is, while cropping the only fancy QR code part from the provided image. We noticed that the python libraries found difficulty to detect the edges (i.e., region of fancy QR code). We tested our algorithm with \textit{pyzbar}, \textit{zxing}, and \textit{cv2} libraries for cropping. Surprisingly, some of the fancy QR codes in our dataset were identified by \textit{cv2} but not all of them and we observed the same experience with other libraries. Therefore, specifically in our app experiments, we selected only those fancy codes that can be detectable by the \textit{cv2} library and for cropping, we developed a manual approach that can successfully crop the fancy codes with white background\par

Also, for the mobile application, we relied on FLASK api server because the pre-trained model and python script to extract features from fancy QR codes are hosted on that server. This required users' device to always connect with an internet service. Additionally, in future, we are planning to implement our approach along with one of the existing scanning approaches such as \cite{su2021artcoder,xu2021art,su2021q,liao2025diffqrcoder}. In these studies, authors proposed various techniques to improve the robustness of scanning process of fancy QR codes. We can determine how this scanning process works specific to identify the QR modules from any fancy QR code. Following this, we then can integrate our methodology which can further extract all the interpretable structural features and pass them to ML model for prediction. In this way, we believe that above highlighted challenges of identifying fancy QR modules and its cropping can be effectively addressed.

\section{Conclusion}
QR codes have become the first point of interaction for individuals in events, restaurants, and online market places, etc, around the world. To secure this interaction, this study presented a safe-by-design approach "ALFA", to cope with Quishing attacks via fancy QR codes. We showed that our approach can effectively identify the irregular representation of fancy QR codes to detect its legitimacy without relying on any deep learning based method or having the need to first classify its potentially malicious payload. We also introduced \textit{FAST} method that can identify and recover any mislabeling of modules according to the standard patterns of QR code. Then, we performed the experimental analysis on diverse examples of fancy QR codes dataset. Furthermore, the real-world applicability of our approach is supported by the developed mobile app and the comparison of its evaluation results with existing real-world QR readers. The overall results from our experiments indicate that this is a timely solution based on the mitigation of Quishing attacks that can also work along within the prevailing QR readers. We also discussed few areas of our work where further improvement is needed. Thereby, we intend to address them in future.

\subsubsection{\discintname}
The authors have no competing interests to declare regarding this paper.



%
%

\bibliographystyle{splncs04}
\bibliography{References}

\end{document}